     \documentstyle[12pt,epsfig]{article}   
     \newlength{\dinwidth}                       
     \newlength{\dinmargin}                      
\def\Journal#1#2#3#4{{#1} {\bf #2}, #3 (#4)}

\def\NPB{{\em Nucl. Phys.} B}
\def\PLB{{\em Phys. Lett.}  B}
\def\PRL{\em Phys. Rev. Lett.}
\def\PRD{{\em Phys. Rev.} D}

\def\lsim{\mathrel{\rlap{\lower4pt\hbox{\hskip1pt$\sim$}}
    \raise1pt\hbox{$<$}}}                
\def\gsim{\mathrel{\rlap{\lower4pt\hbox{\hskip1pt$\sim$}}
    \raise1pt\hbox{$>$}}}                
\newcommand{\la}{\lambda}
\newcommand{\bfp}{\mbox{\boldmath $p$}}
\newcommand{\bfk}{\mbox{\boldmath $k$}}
\begin{document}
\renewcommand{\thefootnote}{\fnsymbol{footnote}}
\begin{flushright}
hep-ph/9907269
\end{flushright}
\vspace*{5mm}
\begin{center}  \begin{Large} \begin{bf}
Single Transverse Spin Asymmetries for\\
Semi-inclusive Pion Production in DIS\footnote{Talk delivered by
F. Murgia at the DESY Workshop ``Polarized Protons at High Energies - 
Accelerator Challenges and Physics Opportunities'',
DESY - Hamburg, 17-20 May 1999.}
\\
  \end{bf}  \end{Large}
  \vspace*{5mm}
  \begin{large}
M. Anselmino$^a$, M. Boglione$^b$ and F. Murgia$^c$
  \end{large}
\end{center}
\begin{center}
$^a$ Dipartimento di Fisica Teorica, Universit\`a di Torino and \\
     INFN, Sezione di Torino, Via P. Giuria 1, 10125 Torino, Italy\\
$^b$ Department of Physics and Astronomy, Vrije Universiteit Amsterdam, \\
     De Boelelaan 1081, 1081 HV Amsterdam, The Netherlands \\
$^c$ INFN, Sezione di Cagliari and Dipartimento di Fisica,
     Universit\`a di Cagliari,\\
     C.P. 170, I-09042 Monserrato (CA), Italy
\end{center}

\setcounter{footnote}{0}
\renewcommand{\thefootnote}{\arabic{footnote}}

\begin{quotation}
\noindent
{\bf Abstract:}
We present a phenomenological approach to the description of single
transverse spin asymmetries (STSA) in inclusive hadron production. It
generalizes the pQCD formalism for high energies and large $p_T$
inclusive hadron production processes, $AB\to C\,X$, by incorporating
quark intrinsic motion in the spin dependent quark distribution
and fragmentation functions. We concentrate here on spin and
$\bfk_\perp$ effects in the fragmentation process of a polarized quark 
into the observed hadron, and give predictions
for STSA in semi-inclusive deep inelastic scattering,
$\ell p^{\uparrow}\to\pi X$ and $\gamma^* p^{\uparrow}\to\pi X$.
\end{quotation}

Polarized inclusive hadron production processes at high energies and $p_T$,
$A,S_A+B,S_B\to C\,X$, can be described in the formalism
of perturbative QCD (pQCD), by using the factorization theorem at leading
twist; the corresponding cross section reads

\begin{eqnarray}
\frac{E_C \, d^3\sigma^{A,S_A + B,S_B \to C + X}} {d^{3} \bfp_C} &=&
\sum_{a,b,c,d;\{\lambda\}}\int \frac {dx_a \, dx_b} {16 \pi^2 z \hat s^2}
\,\rho_{\la^{\,}_a, \la^{\prime}_a}^{a/A,S_A} \, f_{a/A}(x_a) \,
\rho_{\la^{\,}_b, \la^{\prime}_b}^{b/B,S_B} \, f_{b/B}(x_b)
\nonumber\\ &\times&
\hat M_{\la^{\,}_c, \la^{\,}_d; \la^{\,}_a, \la^{\,}_b} \,
\hat M^*_{\la^{\prime}_c, \la^{\,}_d; \la^{\prime}_a, \la^{\prime}_b} \,
D_{\la^{\,}_C,\la^{\,}_C}^{\la^{\,}_c,\la^{\prime}_c}(z) \>,
\label{abcx}
\end{eqnarray}

\noindent
where $S_A$ and $S_B$ specify the polarization of the initial hadrons,
$\{\la\}$ indicates summation over all helicity indices, and
the remaining notation should be self-explanatory.
We will consider here the case of STSA, $A_N = (d\sigma^\uparrow-
d\sigma^\downarrow)/(d\sigma^\uparrow+d\sigma^\downarrow)$, which involves
Eq.~(\ref{abcx}) for $S_A=\uparrow,\downarrow$ and unpolarized hadron $B$;
$\uparrow$,$\downarrow$ refer to spin orientations perpendicular to the
scattering plane ($\uparrow,\downarrow = \pm\,\bfp_A\times\bfp_C/
|\bfp_A\times\bfp_C|$).

It is  however well known that pQCD formalism at leading twist
predicts vanishing STSA, in contrast to recent experimental results
for $p^{\uparrow}(\bar{p}^{\uparrow})\,p\to\pi\,X$
at moderately large c.m. energies and $p_T$ \cite{e704}.
In the last years, thus, a lot of theoretical work has been devoted to reconcile
pQCD predictions on STSA with experimental results, by extending the
theoretical formalism to include higher-twist effects, which can play
a relevant role in the range of $p_T$ values probed at present.
In particular, possible origins of STSA can be introduced by considering
intrinsic, transverse momentum ($\bfk_\perp$), effects in the spin-dependent
quark distribution \cite{sive} and fragmentation \cite{coll} functions.
Following this program, in a series of papers \cite{abm1,abm2} we have
developed a consistent phenomenological approach, assuming that the
pQCD, factorization scheme of Eq.~(\ref{abcx}) holds also when intrinsic
parton motion is included. This way one finds from Eq.~({\ref{abcx}),
at leading order in $\bfk_\perp$, e.g. in the
$p^{\uparrow}\,p\to\pi\,X$ process:

\begin{eqnarray}
& &\!\!\!\!\!\!\!\!\!\! \frac{E_\pi \, d^3\sigma^\uparrow} {d^{3} \bfp_\pi}
- \frac{E_\pi \, d^3\sigma^\downarrow} {d^{3} \bfp_\pi} =
2\ d\sigma^{unp}\ A_N(p^\uparrow p\to\pi X) = 
\sum_{a,b,c,d} \int \frac {dx_a \, dx_b} {\pi z} \nonumber\\
&\times&\,\Bigg\{ \int d^2 \bfk_{\perp} \,
\Delta^Nf_{a/p^\uparrow} (x_A,\bfk_{\perp}) \> f_{b/p}(x_b) \,
\frac{d \hat \sigma} {d\hat t} (x_a,x_b,\bfk_{\perp}) \>
D_{\pi/c}(z) \nonumber\\
&+&\,\int d^2 \bfk'_{\perp}\, h_1^{a/p}(x_a) \>
\Delta^Nf_{b^\uparrow/p} (x_b,\bfk'_{\perp}) \>
\Delta'_{NN} \hat\sigma(x_a,x_b,\bfk'_\perp) \>
D_{\pi/c}(z) \nonumber\\
&+&\,\int d^2 \bfk''_{\perp}\, h_1^{a/p}(x_a) \> f_{b/p}(x_b) \>
\Delta_{NN} \hat\sigma(x_a,x_b,\bfk''_\perp) \>
\Delta^N D_{\pi/c}(z,\bfk''_\perp) \Bigg\} 
\label{akt}
\end{eqnarray}

\noindent
where the first two terms in brackets (respectively
the so-called Sivers effect \cite{sive} and the contribution recently
proposed by Boer \cite{boer}) account for $\bfk_\perp$
effects in the quark distributions inside the initial hadrons
while the last term corresponds to the so-called Collins effect \cite{coll}.
The new, unknown, $\bfk_\perp$ and spin dependent functions of
Eq.~(\ref{akt}) have a simple partonic interpretation as follows:
$\Delta^N f_{q/p^\uparrow}(x,\bfk_\perp) =
\hat f_{q/p^\uparrow}(x, \bfk_{\perp}) -
\hat f_{q/p^\downarrow}(x, \bfk_{\perp})$;
$\Delta^N f_{q^\uparrow/p}(x,\bfk_\perp) =
\hat f_{q^\uparrow/p}(x, \bfk_{\perp}) -
\hat f_{q^\downarrow/p}(x, \bfk_{\perp})$;
$\Delta^N D_{h/q}(z,\bfk_\perp) =
\hat D_{h/q^\uparrow}(z, \bfk_{\perp}) -
\hat D_{h/q^\downarrow}(z, \bfk_{\perp})$.

%
%
%

The other quantities appearing in Eq.~(\ref{akt}), apart from the unpolarized
quark distribution and fragmentation functions, $f$ and $D$, are the
transverse spin content of the proton,
$h_1^{q/p} = f_{q^\uparrow/p^\uparrow}(x) - f_{q^\downarrow/p^\uparrow}(x)$,
and the elementary double spin asymmetries, computable in pQCD:
$\Delta_{NN} \hat\sigma = d\hat \sigma^{a^\uparrow b \to c^\uparrow d}/
d\hat t\,-\,
d\hat \sigma^{a^\uparrow b \to c^\downarrow d}/d\hat t$;
$\Delta'_{NN} \hat\sigma = d\hat \sigma^{a^\uparrow b^\uparrow \to c d}
/d\hat t\,-\,d\hat \sigma^{a^\uparrow b^\downarrow \to c d}/d\hat t$.

%

Notice that the new distributions $\Delta^N f_{q/p^\uparrow}$ and
$\Delta^N f_{q^\uparrow/p}$ are
T-odd functions. In order not to vanish they require some
initial-state interactions (with possible breaking of factorization
theorem and universality), or finite-size effects in the proton,
or spin-isospin interactions.

Eq.~(\ref{akt}) may be used to obtain information on the new,
$\bfk_\perp$-dependent distributions, by fitting experimental data
on STSA in inclusive hadron production.
This program has been started and developed in \cite{abm1},
where Sivers effect was considered
as the only possible source for STSA, and continued in \cite{abm2}, where
a similar study was performed in the case of Collins effect alone.
It was shown that both effects are separately able to reproduce well the
experimental data on $p^\uparrow p\to\pi X$. However, Collins effect
alone appears to have some problems, since a reasonable fit is obtained only
at the price of saturating the natural positivity bound
$|\Delta^N D_{\pi/q}(z)| \leq 2 D_{\pi/q}(z)$
at large $z$ values.
The Soffer's inequality for the transversity distribution,
$2|h_1^{q/p}(x)|\le q(x)+\Delta q(x)$, could also be violated
by the parametrization of the fit\footnote{We are grateful to E. Leader
for drawing our attention on this point.}.

Using the information on $\Delta^N D_{h/q}(z,\bfk_\perp)$ obtained in
\cite{abm2}, we apply here our approach to give predictions for STSA in the
semi-inclusive pion production in DIS. These processes are particularly
interesting because we expect that, since Sivers
effect requires initial state interactions which are suppressed by
$\alpha_{em}$ in $\ell p$ interactions, Collins effect should be entirely
responsible for STSA. An experimental study of these processes should then be
very useful for testing the relevance of Collins effect in explaining STSA.
A more detailed treatment can be found in \cite{abhm}.

We first consider the process $\ell p^\uparrow\to\pi X$. If Collins effect
is the dominant effect responsible for the sizeable STSA observed in
the analogous process $p^\uparrow p\to\pi X$, we expect large asymmetries in
this case too. Using Eq.~(\ref{akt}), we get

\begin{equation}
2\ d\sigma^{unp}\ A_N(\ell p^\uparrow\to\pi X) =
\sum_q \int \frac {dx} {\pi z}
\> \int d^2 \bfk_{\perp}\,
\left [ h_1^{q/p}(x) \> \Delta_{NN} \hat\sigma^q(x, \bfk_\perp) \>
\Delta^N D_{\pi/q}(z, \bfk_\perp) \right ] 
\label{gendis}
\end{equation}

\noindent
where $\Delta_{NN} \hat\sigma^q = d\hat \sigma^{\ell q^\uparrow \to
\ell q^\uparrow}/d\hat t\ -\ 
d\hat \sigma^{\ell q^\uparrow \to \ell q^\downarrow}/d\hat t$.


As an example, in Fig. 1 the STSA $A_N(\ell p^\uparrow\to\pi X)$ is shown
for typical HERMES kinematics, as a function of $x_F$. Similar results
hold for different kinematical situations corresponding to other available
experimental set-ups \cite{abhm}.

\begin{figure}[t]
\begin{center}
\epsfig{figure=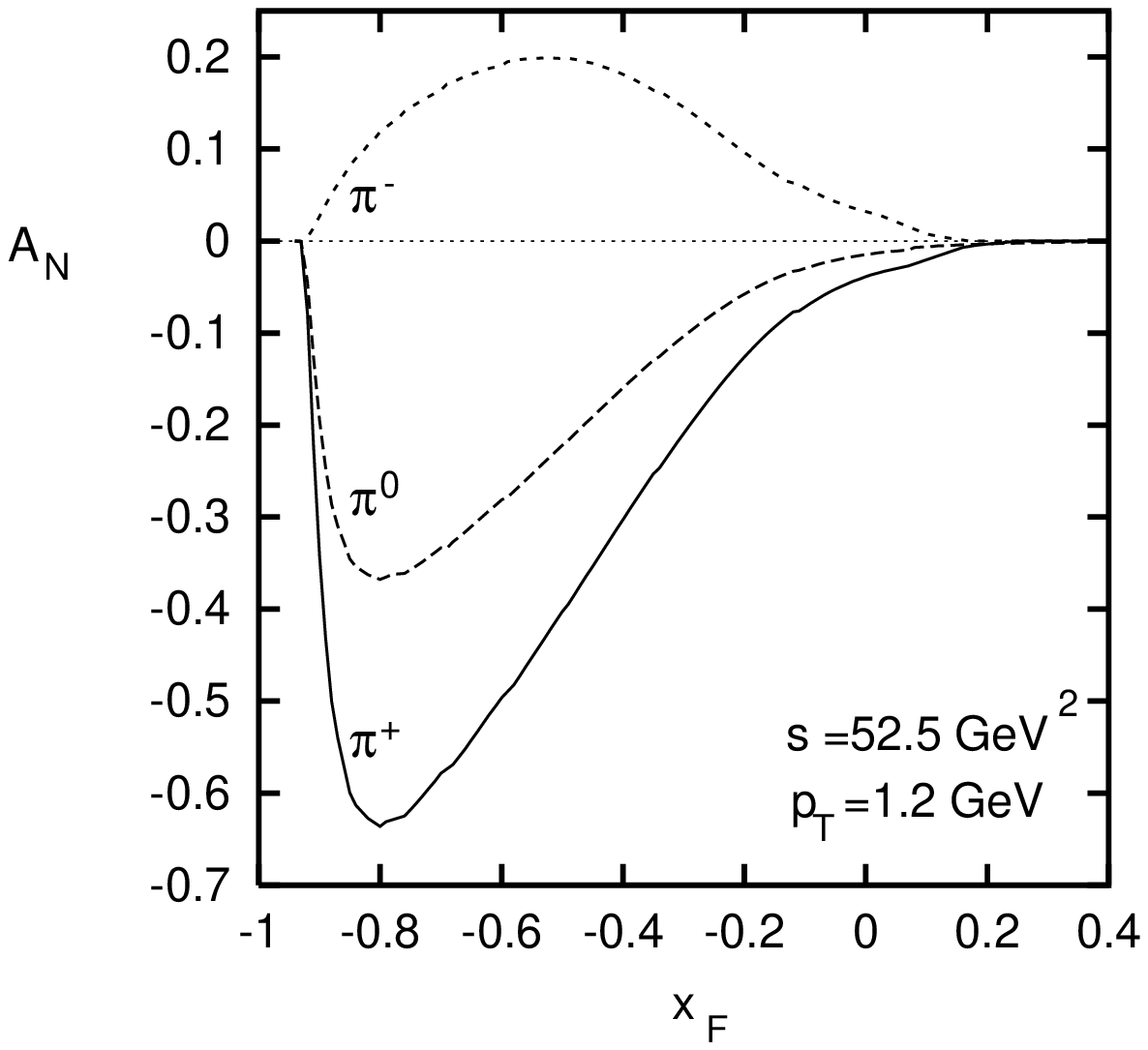,angle=-90,width=10cm}
 \begin{minipage}[c]{10cm}
\vspace{16pt}
\baselineskip=3pt
 {\footnotesize {\bf Fig. 1:}
  $A_N(\ell p^{\uparrow}\to\pi X)$ for typical HERMES kinematics.}
 \end{minipage}
 \end{center}
\vspace{-13pt}
\end{figure}

A measurement of STSA in $\ell p^\uparrow\to\pi X$ requires transversely
polarized nucleons, which are presently available only for some experiments.
However, STSA may be measurable also in the case of longitudinally
polarized nucleons, provided one looks at the double inclusive process,
$\ell p^\uparrow \to \ell \pi X$ from which one can reconstruct
the $\gamma^* p^\uparrow \to \pi X$ reaction, which, in general, occurs in
a plane different from the $\ell-\ell'$ plane where the longitudinal
nucleon spin lies.
In this case one has (see \cite{abhm} for details):
\begin{equation}
\frac{d\sigma^{\gamma^* p^\uparrow \to \pi X}}{dx \, dQ^2 \, dz \, d^2p_T}
- \frac{d\sigma^{\gamma^* p^\downarrow \to \pi X}}{dx \, dQ^2 \, dz \, d^2p_T}
= \sum_q\,
h_1^{q/p^\uparrow}
\left[ \frac{d\hat\sigma^{\gamma^* q^\uparrow \to q^\uparrow}} {dQ^2}
- \frac{d\hat\sigma^{\gamma^* q^\uparrow \to q^\downarrow}} {dQ^2} \right] \>
\Delta^ND_{\pi/q}(p_T) \>.
\label{ancg}
\end{equation}

In Fig.~2 we show $A_N$ at the same energy values of Fig.~1, as
a function of $z$: it is large also in this case, although only at very
large $z$ values which might be difficult to reach experimentally.

In conclusion, let us remark that semi-inclusive hadron production in
DIS may be crucial for testing the relevance of Collins effect in
explaining sizeable STSA.
In this context, HERMES and future planned spin experiments at HERA
can surely play a fundamental role.

\begin{figure}[t]
\begin{center}
\epsfig{figure=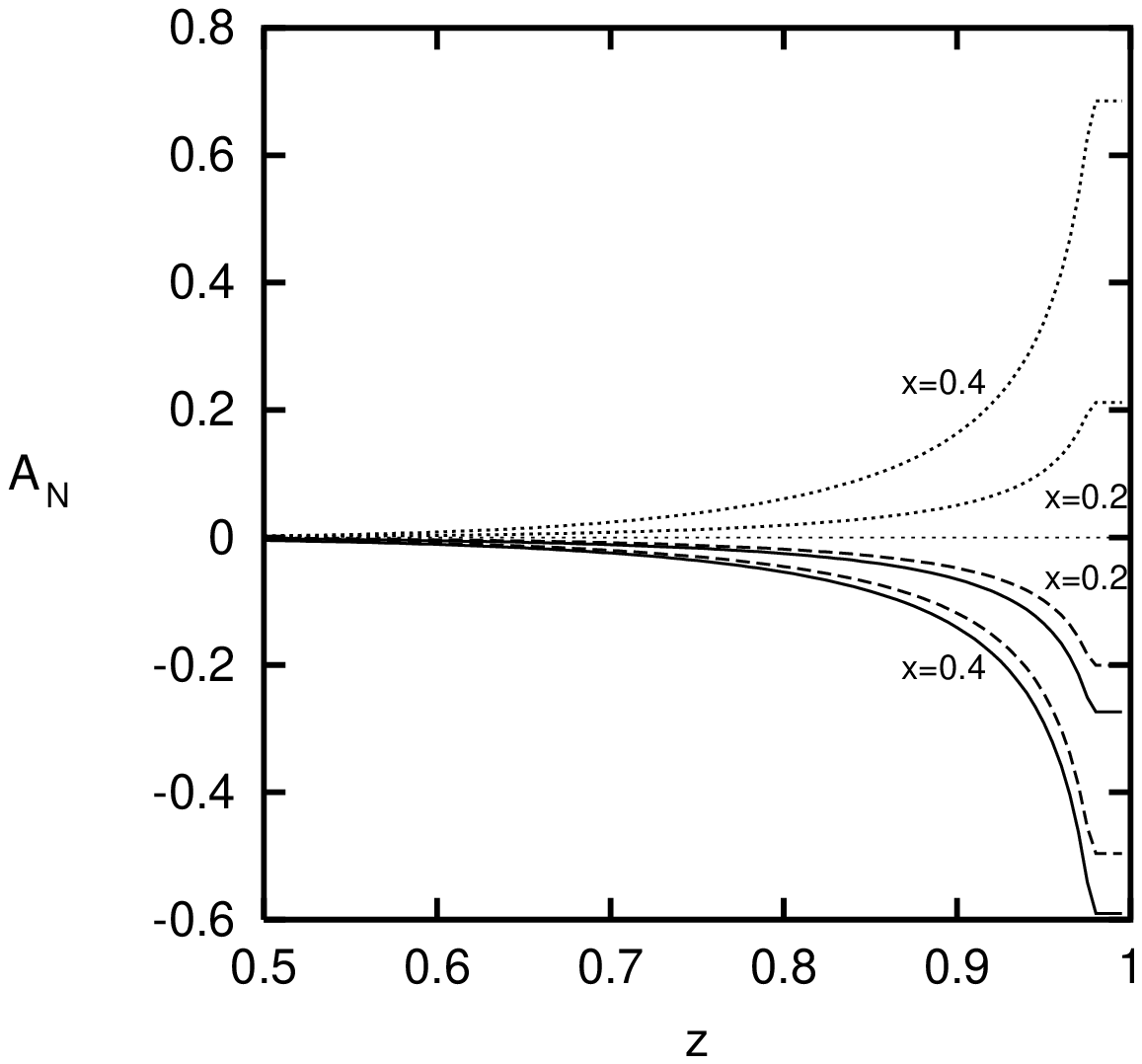,angle=-90,width=10cm}
 \begin{minipage}[c]{10cm}
\vspace{12pt}
\baselineskip=3pt
 {\footnotesize {\bf Fig. 2:}
 $A_N(\gamma^*p^{\uparrow}\to\pi X)$: $s=52.5$ GeV$^2$, $Q^2 = 8$
 GeV$^2$; solid, dashed and dotted line refer respectively to $\pi^+$, $\pi^0$
 and $\pi^-$.} 
 \end{minipage}
 \end{center}
\vspace{-13pt}
\end{figure}

\vspace{8pt}

\noindent
{\Large\bf Acknowledgements}
\vspace{6pt}

\noindent
One of us (FM) wishes to thank the organizers of this very interesting
workshop for the warm hospitality.

\end{document}